\documentclass[aps,prl,twocolumn,groupedaddress,showpacs,amsmath,
amssymb,amsfonts]{revtex4}
\usepackage{amsmath}
\usepackage{amsfonts}
\usepackage{amssymb}
\usepackage{bm}
\usepackage{graphicx}%
\def\eps{\varepsilon}

\begin{document}
\title{Spot deformation and replication in the two-dimensional
Belousov-Zhabotinski reaction in water-in-oil microemulsion}
\author{Theodore Kolokolnikov}
\author{Mustapha Tlidi}
\email[]{tkolokol@mathstat.dal.ca}
\email[]{mtlidi@ulb.ac.be}
\affiliation{Department of Mathematics and Statistics, Dalhousie
University, Halifax, Canada}
\affiliation{Optique Nonlin\'eaire Th\'eorique, Universit\'e Libre de Bruxelles, Campus Plaine CP 231, 
1050 Bruxelles, Belgium}

\date{\today}
\begin{abstract}
In the limit of large diffusivity ratio, 
spot-like solutions in the two-dimensional Belousov-Zhabotinski 
reaction in water-in-oil microemulsion
are studied. It is
shown analytically that
such spots undergo an instability as the diffusivity ratio is
decreased. An instability threshold is derived. For spots of 
small radius, it is shown that this instability leads to a spot splitting
into precisely two spots. For larger spots, it leads to deformation,
fingering patterns and space-filling curves.
Numerical simulations are shown to be in close agreement with the
analytical predictions.
\end{abstract}


\maketitle
Localized patterns such as spots belong to the class of dissipative
structures found far from equilibrium  \cite{DStruc}.  In recent years,
considerable progress has been made in the understanding of these
systems. 
The question of stability of such patterns is central and the source of
instabilities must be carefully scrutinized.  In particular, the
occurrence of instability can lead to the deformation of spots followed
by spot multiplication (also called self-replication) or fingering.  This
intriguing phenomenon has been the subject of research since the
pioneering work of Pearson  \cite{pearson}.  Shortly after, thanks to the
development of open spatial chemical reactors, self-replication was
observed in various experiments such as ferrocyanide-iodate-sulphite
reaction \cite{lmps}, the Belousov-Zhabotinsky reaction \cite{mpm, kve,
kve2},
and chloride dioxide-malonic-acid reaction \cite{DBDK}. By now this
phenomenon is believed to be universal \cite{mo, hayase1}. It is not
restricted to chemical reactions and occurs in many 
systems in biology \cite{meron}, material science \cite{RW, ns} and
nonlinear optics \cite{laser}. 

Analytically, spot replication is relatively well understood in
one dimensional setting. Nishiura and Ueyama \cite{N5} proposed that
self-replication of spikes occurs when the spike solution dissapears due
to the presence of a fold point.  Similar explanation has been reported
for the box-like patterns \cite{OK}.  In two dimensions, 
a mechanism for spot instability has been
proposed in \cite{mo} for general reaction-diffusion systems; related
analysis was performed earlier in the framework of  a piecewise-linear
approximation in \cite{omk}, \cite{km} and more recently in the context
of diblock copolymer systems \cite{RW}, \cite{ns}. 
See also \cite{Goldstein}, \cite{Mmotion} and \cite{gcom} and for a related 
approach from the point of view of interface motion.

In this letter we perform an analytical and numerical investigation of
the two-dimensional localized spots that were recently reported for the
Belousov-Zhabotinski (BZ) reaction in water-in-oil microemulsion
\cite{kve}, \cite{kve2}.
In the classical BZ reaction, spiral waves are observed \cite{KT} but no
localized spot solutions are possible.  Indeed, localized structures
develop only when the ratio of diffusion coefficients is sufficiently
large, which occurs in the microemulsion system but not in the
classical BZ reaction.

We consider the
water-in-oil microemulsion model of the BZ reaction as described in 
\cite{kve} and \cite{kve2}: 
\begin{subequations}
\label{bz}
\begin{align}
&\begin{aligned}
\varepsilon_{0}v_{t}  = 
\varepsilon_{0}D_{v}\Delta v+\left[  f_{0}
z+i_{0}\left(  1-mz\right)  \right]  \frac{v-q_{0}}{v+q_{0}}\\
+\left[
\frac{1-mz}{1-mz+\varepsilon_{1}}\right]  v-v^{2}
\end{aligned}\\
&z_{t}    =D_{z}\Delta z-z+v\left[  \frac{1-mz}{1-mz+\varepsilon_{1}}\right]
\end{align}
\end{subequations}
where $v, z$ are dimensionless concentrations of activator HBrO$_2$ and
oxidized catalyst $[Ru(bpy)_3]^{3+}$ respectively; $D_v$ and $D_z$ are
dimensionless diffusion coefficients of activator and catalyst; $f, \eps$
and $q$ are parameters of the standard Keener-Tyson model \cite{KT}; $i_0$
represents the photoinduced production of inhibitor, and $m$ represents
the strength of oxidized state of the catalyst with $0 <mz < 1$.
This reaction was shown experimentally and
numerically to admit localized spot patterns that persist for long time
\cite{kve}, \cite{kve2}. 

We rescale the variables as
$
z=1/m-m^{-3/2}w\varepsilon_{1},~~v=m^{-1/2}\hat{v},~~ t=\varepsilon
_{0}m^{1/2}\hat{t}.
$
In the new variables, after dropping the hats, we obtain
\begin{subequations}
\begin{align}
v_{t} =\varepsilon^{2}\Delta v+f(v,w);\ \ \tau w_{t}=D\Delta
w+g(v,w) \label{vw}
\end{align}
where
\begin{align}
f(v,z)   &=-\left[  f_{0}+f_{1}w\right]  \frac{v-q}{v+q}+\left[
\frac{w}{1+\alpha w}\right]  v-v^{2};\\
g(v,w)&=1-\left[  \frac
{w}{1+\alpha w}\right]  v
\end{align}
\end{subequations}
and with the nondimensional constants given by
\begin{align}
\begin{aligned}
&\alpha   =m^{-1/2},~~ f_{1}=\varepsilon_{1}m^{1/2}\left(  i_{0}%
-\frac{f_{0}}{m}\right),~~
q  =q_{0}m^{1/2},\\
&\varepsilon^2=\varepsilon_{0}D_{v}m^{1/2}
,~~ D=D_{z}\varepsilon_{1}m^{-1/2},~~ \tau=\frac{1}{m}\frac{\varepsilon_{1}
}{\varepsilon_{0}}.
\end{aligned}
\end{align}
In the limit $m\rightarrow\infty,\ f_{1}\rightarrow0$ and $\tau\rightarrow
0\ \ $we obtain the reduced system,%
\begin{equation}
\left\{
\begin{aligned}
v_{t}&=\varepsilon^{2}\Delta v-f_{0}\frac{v-q}{v+q}+wv-v^{2}\\
0&=D\Delta w+1-vw.
\end{aligned}
\right.  \label{simple}%
\end{equation}
In particular, parameter values used in Fig.~14 of \cite{kve2} are 
$f_0=2.18, i_0=0,
m=10, \eps_1 = 0.01, \eps_0=0.1$ and $D_z/D_v=100$ which gives $\alpha=0.3,
f_1 = -0.007, \tau=0.01$, 
so that the simplification (\ref{simple}) is appropriate.

{ \begin{figure}[b]
{ \begin{minipage}[t]{0.15\textwidth}
\begin{center}{\small
\setlength{\unitlength}{1\textwidth} \begin{picture}(1,0.42)(0,0)
\put(-0.2,0.0){\includegraphics[width=1.2\textwidth,
height=1.1\textwidth]{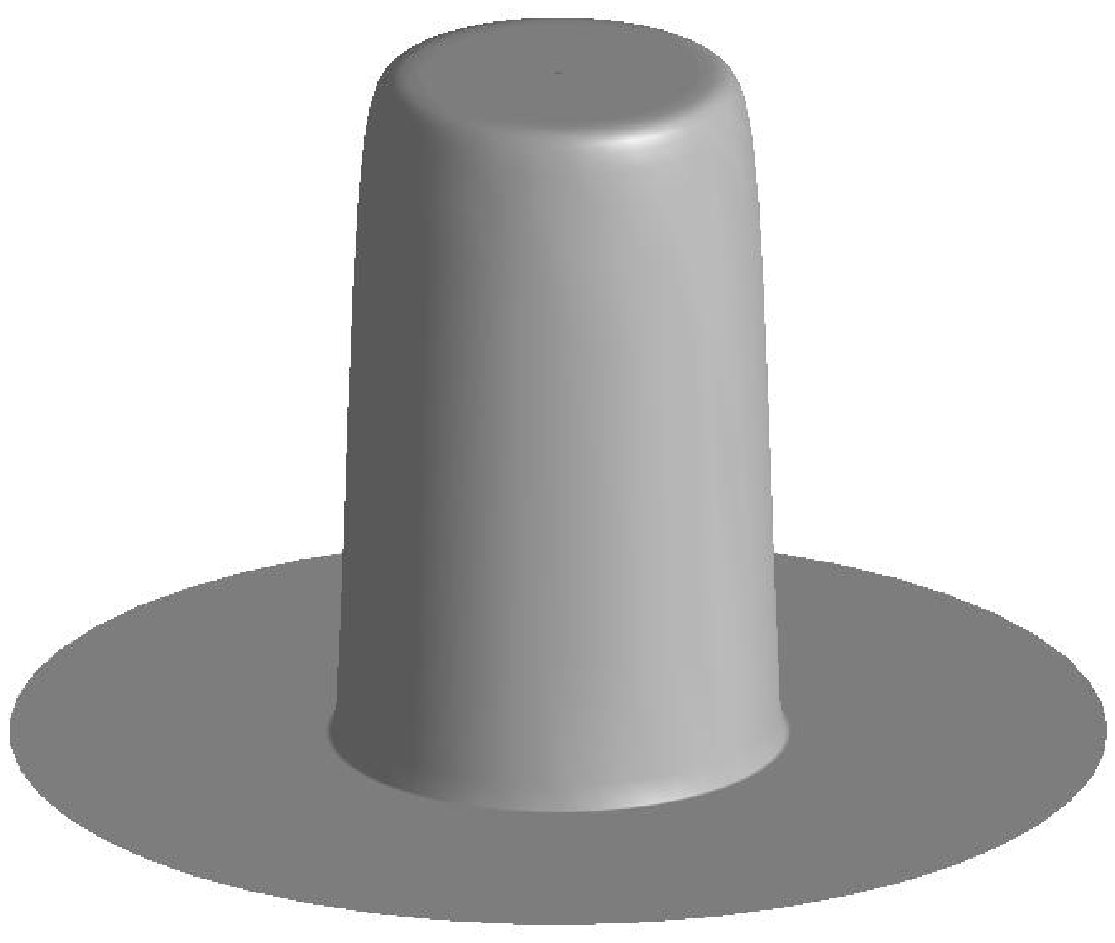}}
\end{picture}
}(a)\end{center}
\end{minipage}
\begin{minipage}[t]{0.30\textwidth}
\begin{center}{
\setlength{\unitlength}{1\textwidth} \begin{picture}(1.0,0.42)(0,0)
\put(-0.05,0.03){\includegraphics[height=0.42\textwidth,
]{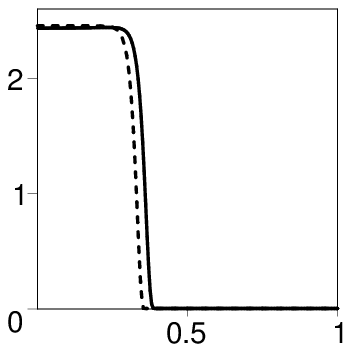}}
\put(+0.21,0.28){\includegraphics[height=0.12\textwidth,
]{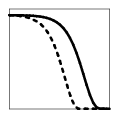}}
\put(0.5,0.03){\includegraphics[height=0.42\textwidth,
]{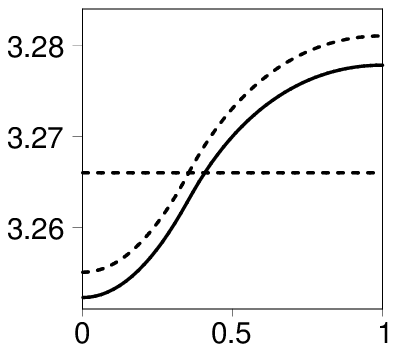}}
\end{picture}
}
(b) \ \ \ \ \ \ \ \ \ \ \ \ \ \ \ \ \ \ \ \ \ \ \  (c)
\end{center}
\end{minipage}
}\caption{(a) Radially symmetric, stable spot solution of the reduced BZ
system (\ref{simple})
in the BZ system on a unit disk.
(b) Radial profile of $v$ 
(solid curve) and its asymptotic approximation (dashed 
curves) given by (\ref{ss}). Insert: a blowup showing the profile of the
interface.
(c) Radial profile of $w$ (solid curve) and its two-term
asympotic approximations
(dashed curves) given by (\ref{w}).
Parameter values are $D = 20, f_0 = 2.0, \eps=0.02, q=0.005.$
}
\label{fig:steady}
\end{figure}} 

Experimental and numerical evidence in \cite{kve} and \cite{kve2} 
suggest that (\ref{bz}) admits localized spot soltuions,
such as shown in Fig.~\ref{fig:steady}.
Such spots occur in the regime where $\eps \ll 1.$
Our goal is to describe analytically the radius and profile of such a spot
and then study its
stability using singular perturbation techniques 
similar to those described in \cite{mo}.
As we will demonstrate, the instability thresholds appear in
the regime where  $D \gg 1$.  For sufficiently large values of $D$ a
spot pattern is stable. However as $D$ is decreased, instabilities of the
form $\exp(\lambda t) \cos(m\theta) \phi(r)$ may develop, where $\theta$
and $r$ are the angular and radial coordinates, respectively.
We provide the analytic description of the profile of the spot
(\ref{ss}, \ref{w}) and the dispersion relation (\ref{main-result}) 
between $m$ and $\lambda$. This relation leads directly to the estimate
for the instability threshold.  

We begin by constructing a stationary (time-independent)\ spot-type solution
on a two dimensional unit disk $\left\{  x:\left\vert x\right\vert <1\right\}
$. We assume that $D\gg1$. Then to leading order, $w\sim w_{0}$
is a constant to be determined, and the solution for $v$ consists of an
interface located at $\left\vert x\right\vert \sim l$ which connects two
nearly spatially homogeneous layers. To find the profile of such an interface
and its location $l$, let us rescale near $l$ as $v\left(  x\right)  =V\left(
y\right)  $ with $y=\left(  r-l\right)  /\varepsilon$ and $r=\left\vert
x\right\vert .$ For the steady state we then obtain, to leading order,
$V^{\prime\prime}\left(  y\right)  +f\left(  V;w_{0}\right)  =0.$ The
interface solution corresponds to a heteroclinic orbit of this ODE. The
existence of such an orbit is only possible whenever the equations
\begin{equation}
f\left(  V_{-};z_{0}\right)  =0=f\left(  V_{+};w_{0}\right)  ,\ \ \ \ \int
_{V_{-}}^{V_{+}}f\left(  s;w_{0}\right)  ds=0\label{Vpm}%
\end{equation}
are simultaneously satisfied for some values $V_{-}$ $<V_{+}.$ Since $q\ll1$,
we have $V_{-}\sim 0$ and (\ref{Vpm}) can be written
as
\begin{equation}
f_{0}+f_{1}w_{0}\sim\frac{3}{16}\left[  \frac{w_{0}}{1+\alpha w_{0}}\right]
^{2};\ \ \ V_{+}\sim\frac{3}{4}\left[  \frac{w_{0}}{1+\alpha w_{0}}\right]
.\label{Vpw0}%
\end{equation}
Next, we ignore the $O(q)$ terms and integrate $V''(y)+f(V,w_0)=0$ 
to obtain
\begin{equation}
\begin{aligned}
v&\sim\left\{
\begin{aligned}
&V_{+}\tanh^{2}\left(  \sqrt{\frac{V_{+}}{6}}\left(  \dfrac{r-l}{\varepsilon
}\right)  \right)  ,~~& r<l\\
&0,& r>0
\end{aligned}
\right.  
\end{aligned}
\label{ss}%
\end{equation}
with $V_{+}$ and $w_{0}$ given by (\ref{Vpw0}) and where $r=\left\vert
x\right\vert $. This formula describes the profile of the interface in
the limit $q \to 0$. Its
thickness is of $O(\eps V_+^{-1/2})  .$ To determine its location
$l$, we integrate the second equation in (\ref{vw}). Zero-flux conditions then
yield $\int g=0$ so that
\begin{equation}
g\left(  V_{+},w_{0}\right)  l^{2}+\left(  1-l^{2}\right)  g\left(
0,w_{0}\right)  \sim0\label{gio}%
\end{equation}
In the limit of the reduced system (\ref{simple}),\ we obtain%
\begin{align}
\label{ss2}
l\sim\frac{1}{2\sqrt{f_{0}}},\ \ w_{0}\sim\frac{4}{\sqrt{3}}\sqrt{f_{0}%
},\ \ \ V_{+}\sim\sqrt{3f_{0}}.
\end{align}
To determine the correction to $w,$ we write $w=w_{0}+D^{-1}%
w_{1}+O\left(  D^{-2}\right)  $. to obtain $\Delta w_{1}+g\left(  w_{0}%
,v_{0}\right)  =0.$ Imposing continuity at the interface $r=l$, the
solvability condition $w_{1}(l)=0$, and using (\ref{gio})\ and $g\left(
0,w_{0}\right)  =1,$ we then obtain
\begin{align}
w\sim w_{0}+\frac{1}{4D}\left\{
\begin{aligned}
&-\dfrac{(1-l^2)(l^2-r^2)}{l^2}, &r<l\\
&
2\ln(\dfrac{r}{l})+l^2-r^2, &r>l
\end{aligned}
\right. .
\label{w}
\end{align}
An example of a localized spot and its radial profile are shown in Fig 
\ref{fig:steady}.

\begin{figure}[tb]
\setlength{\unitlength}{0.5\textwidth} \begin{picture}(1,0.30)(0,0)
\put(0.0,0.17){\footnotesize $\dfrac{\lambda}{\eps^2}$}
\put(0.04,0.09){\includegraphics[width=0.12\textwidth,
height=0.08\textheight]{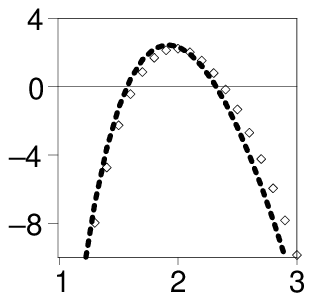}}
\put(0.18,0.056){\footnotesize $m$}
\put(0.18,0.00){(a)}
\put(0.38,0.09){\includegraphics[width=0.12\textwidth,
height=0.08\textheight]{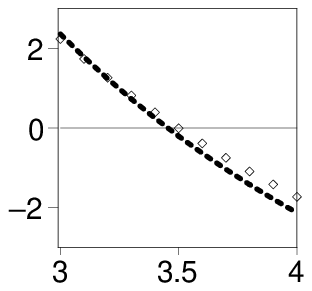}}
\put(0.33,0.17){\footnotesize $\dfrac{\lambda}{\eps^2}$}
\put(0.51,0.056){\footnotesize $D$}
\put(0.51,0.00){(b)}
\put(0.71,0.09){\includegraphics[width=0.12\textwidth,
height=0.08\textheight]{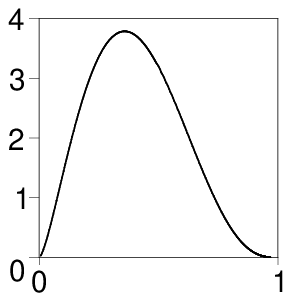}}
\put(0.66,0.17){\footnotesize $D$}
\put(0.79,0.063){\footnotesize $l=\frac{1}{4f_0^2}$}
\put(0.84,0.00){(c)}
\end{picture}
\caption{(a), (b) Comparison of numerical computations of $\lambda$ given by
(\ref{phipsi}) (diamonds) with
the analytical result (\ref{main-result}, \ref{simple-A}) (dashed line)
for the reduced model
(\ref{simple}) on a unit disk. Parameter values are $q=0.01$, $\eps=0.01$ and 
(a) $D=3, f_0=1.3$; (b) $m=2, f_0=1.3$.
(c) Simultaneous solution of $\lambda=0=d\lambda/dm$, showing the first
value of $D$ for which instability occurs. The system is stable above the
curve and unstable below it.
To compute $\lambda$ numerically, 
(\ref{phipsi}) was reformulated as a boundary value problem by adjoining
the equation $d\lambda/dr=0$ along with fixing $\psi(1).$
Maple's numerical 
boundary value problem solver was then used with initial guesses 
$\phi=v_r, \lambda=0$ and $\psi =$ the solution of
(\ref{reducedpsi}).
All computations are
correct to four significant digits. }
\label{fig:evals}
\end{figure}

When decreasing the diffusion coefficient $D$ of the recovery variable,
numerical simulations show that the spot becomes unstable. To compute the
threshold associated with this instability, we linearlize around the
localized spot solution (\ref{ss}) as
\begin{align*}
v(x,t)  &  =v(x)+\exp\left(  \lambda t\right)  \cos\left(  m\theta\right)
\phi\left(  r\right) \\
w(x,t)  &  =w(x)+\exp\left(  \lambda t\right)  \cos\left(  m\theta\right)
\psi\left(  r\right)
\end{align*}
where $v$ and $w$ are given to leading order in (\ref{ss} and \ref{ss2}) spot-type solution, $m$ is an integer, $\phi,\psi\ll1$
and $(r,\theta)$ are the polar coordinates of $x.$ Substituting into
(\ref{bz}) we then obtain
{
\begin{subequations}
\begin{align}\label{phi}
\lambda\phi &  =\varepsilon^{2}\left(  \phi_{rr}+\frac{1}{r}\phi_{r}%
-\frac{m^{2}}{r^{2}}\phi\right)  +f_{v}  \phi+f_{w}  \psi\\ 
\label{psi}
\tau\lambda\psi &  =D\left(  \psi_{rr}+\frac{1}{r}\psi_{r}-\frac{m^{2}}{r^{2}%
}\psi\right)  +g_{v}  \phi+g_{w}  \psi.
\end{align}
\label{phipsi}
\end{subequations} }
Note that $v_r$ satsifies
that $\varepsilon^{2}(v_{rrr}+1/rv_{rr}-1/r^2v_r)+f_{v}v_{r}+w_{r}f_{w}=0.$ 
Therefore multiplying (\ref{psi}) by $v_r$ and integrating by parts, we obtain
\begin{align}\label{2:56}
\left(  \lambda+\frac{(m^{2}-1)\varepsilon^{2}}{l^{2}}\right)  \int v_{r}^{2}%
\sim\int v_{r}f_{w}\left(  \psi-w_{r}\right)  .
\end{align}
where we have assumed that to leading order, 
$\phi\sim v_{r}, \lambda \ll 1, \psi \ll \phi$.
Since $v_{r}$ is exponentially small outside the interface, we simplify $\int
v_{r}f_{w}\left(  \psi-w_{r}\right)  \sim-\left(  \psi\left(  l\right)
-w_{r}\left(  l\right)  \right)  \int_{0}^{V_{+\ }}f_{w}\left(  v,w_{0}%
\right)  dv$. We estimate
$w_r(l)  \sim -1/(2D)g\left(  V_{+},w_{0}\right)$ and
to determine $\psi\left(  l\right),$ we integrate (\ref{psi})
over the interface $l$ to obtain
$
D\psi_{r}|_{l^{-}}^{l^{+}}
\sim-g\left(  0,w_{0}\right)  /l^{2}
$
where we have used $\phi \sim v_r$ and (\ref{gio}). Keeping only
leading order terms in $D$ we then obtain the following problem for
$\psi$:
\begin{subequations}
\label{reducedpsi}
\begin{align}\label{djump}
&0=
\psi_{rr}+\frac{1}{r}\psi_{r}-\frac{m^{2}}{r^{2}}\psi=0,
~~~~ r\ne l
\\
&  \psi^{\prime}(  0)=0=\psi^{\prime}(  1)
,~~
  \psi^{\prime}\left(  l^{+}\right)  -\psi^{\prime}\left(  l^{-}\right)
=-\frac{1}{Dl^{2}}g\left(  0,w_{0}\right)  . 
\end{align}
\end{subequations}
Using the continuity of $\psi$ at $r=l$ we get
\[
\psi\left(  l\right)  \sim\frac{1}{2Dml}g\left(  0,w_{0}\right)  \left(  1+l^{2m}\right)
\]
and substituting into (\ref{2:56}) we obtain

\begin{align}
\lambda\frac{l^{2}}{\varepsilon^{2}}\sim 1-m^{2}+A\left[  1-l^{2}-\frac{1}%
{m}\left(  1+l^{2m}\right)  \right]
\label{main-result}
\end{align}
where
\begin{align}
A=\frac
{l}{2\varepsilon D}\frac{g\left(  0,w_{0}\right)  \left(  \int_{0}^{V_{+\ }%
}f_{w}\left(  v,w_{0}\right)  dv\right)  }{\varepsilon\int v_{r}^{2}}.
\end{align}
For the reduced model (\ref{simple}) we obtain an explicit result
\begin{align}
\label{simple-A}
A \sim \frac{3^{5/4}2^{1/2}5}{64}\frac{1}{f_0^{3/4}\eps D}
=
0.43622\frac{1}{f_{0}^{3/4}\varepsilon D}.
\end{align}
In Fig.~\ref{fig:evals}, the analytical prediction given by
(\ref{main-result}) is found to be in good agreement with the numerical
computations of problem (\ref{phipsi}).  Full numerical simulations of
(\ref{simple}) also agree with this prediction.  For example, when taking
$\eps=0.03, f_0=1.3$ and $q=0.01$, slight spot deformation corresponding
to mode $m=2$ is observed when $D = 1.0$ 
but not when $D=1.3.$ This agrees well with $D\sim1.1$, 
the threshold predicted by (\ref{main-result}).

From (\ref{main-result}) and (\ref{simple-A}), it is clear that 
the mode $m=1$ is always stable and that
for large enough $D$, all modes $m\geq1$ are stable. As $D$ is decreased,
instability sets in when $D=O(\eps^{-1})$.
The threshold
value is found by simultaneously solving $\lambda=d\lambda/dm=0$. 
The resulting graph is shown on Fig.~\ref{fig:evals}c.
In particular, note that the system is stable if $D\eps > 0.038$, independent
of the value of $f_0.$
In the limit
of small radius $l\rightarrow0$, the first unstable integer mode is $m=2$
corresponding to $A=6,$ so that
the spot of small radius becomes unstable whenever 
$\eps D f^{3/4} \le 0.0727.$
More generally, by eliminating $A$ 
we find
that there exist constants $l_{1}<l_{2}<\ldots<1$ such the first unstable mode
is $m$ provided that $l_{m-1}<l<l_{m},$ where $l_1=0$, $l_2=0.491$, $l_3=0.667$,
$l_4=0.753$ and
$l_{m}\sim 1-0.937/m$ as $m\rightarrow\infty;$ where 0.937 is the root of
$e^{-2z}\left(  3+2z\right)  +3-4z=0.$

\begin{figure*}
\begin{center}
\includegraphics[width=0.45\textwidth]{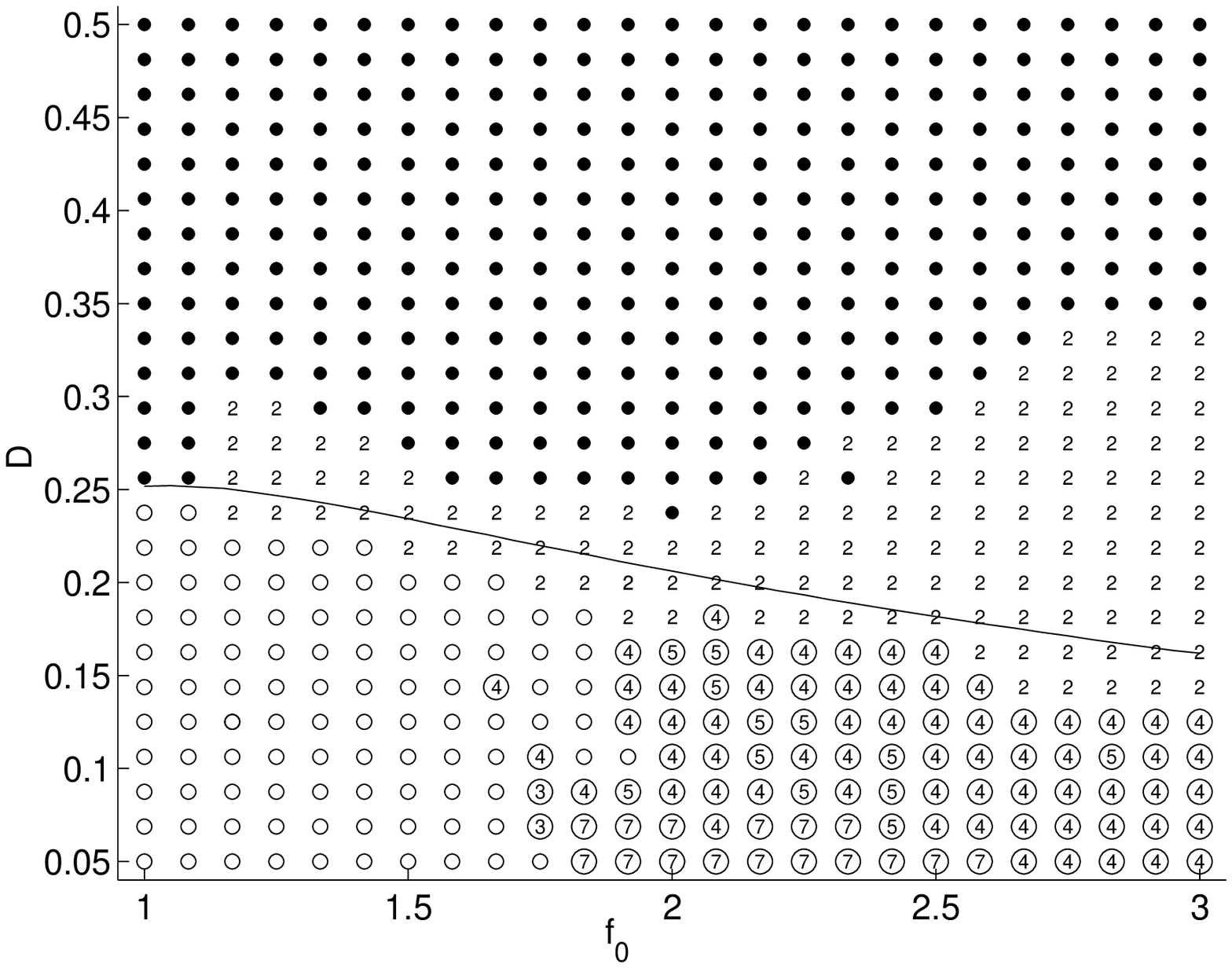}
\includegraphics[width=0.45\textwidth]{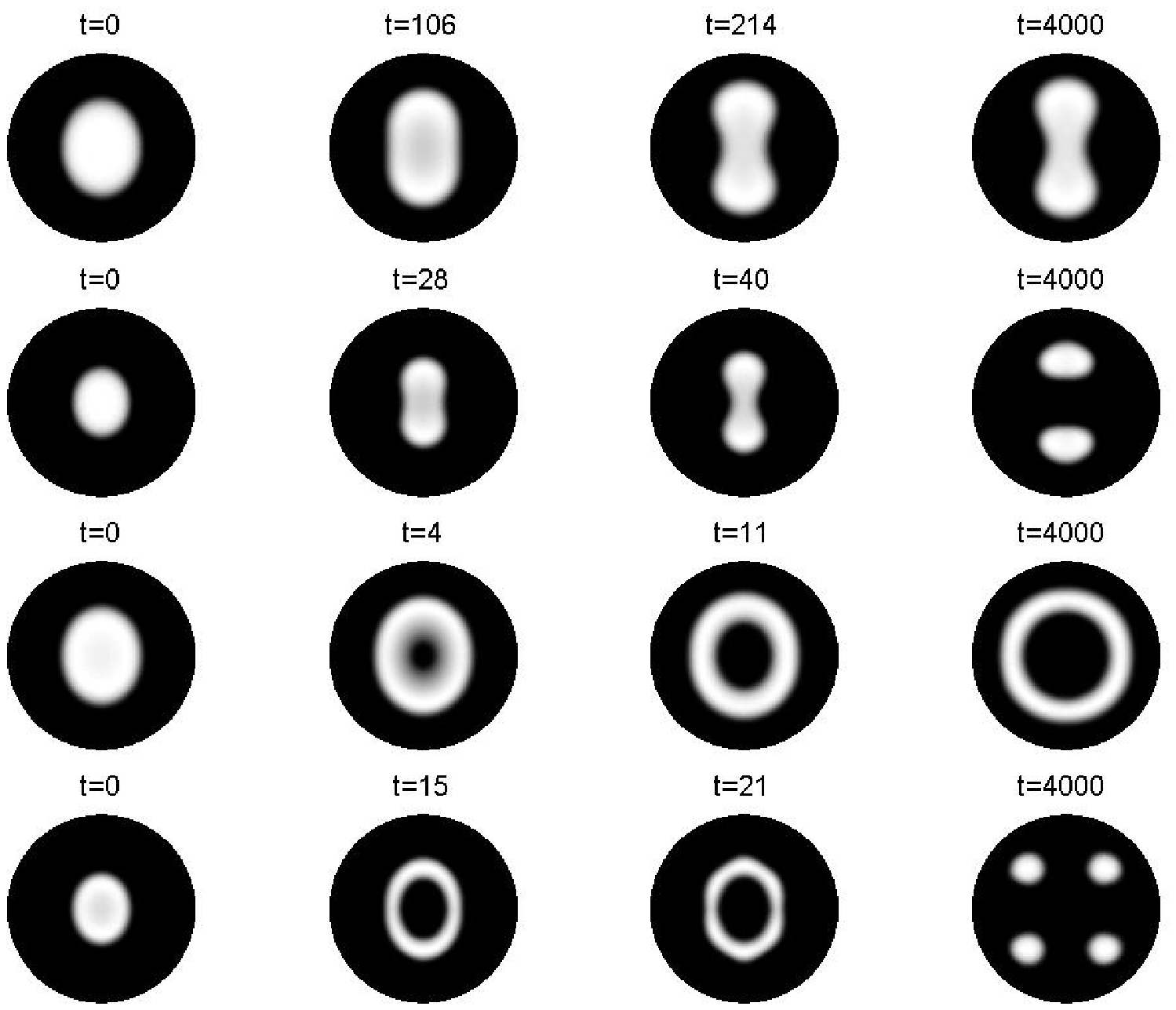}\\
\hfill (a)\hfill$~$\hfill  (b) \hfill$~$\\
\end{center}
\caption{ (a) Bifurcation diagram for (\ref{simple}) in $D$ and $f_0$ with 
$\eps=0.05, q=0.01.$ Solid dots represent deformations of a spot without
topological change. Points marked by "2" represent spot-replication into
two spots. An empty circle represents spot-to-ring bifurcation 
and empty circle
with a number inside represents spot-to-ring-to-spots bifurcation.
A solid line represents the boundary of spot-to-ring replication which
occurs at the fold point of the radially symmetric steady state. 
The bifurcation diagram was obtained by solving the full two-dimensional
system (\ref{simple}) 
using the finite element package {\tt FlexPDE} \cite{FlexPDE} with zero-flux
boundary conditions and 800 elements on a quarter-disk. For initial
conditions, (\ref{ss}, \ref{ss2}, \ref{w}) was used, but with $r$
replaced by $r(1+0.05 \cos2\theta)$.
The solid
line was obtained by solving for the fold point of the radially symmetric
steady state using Maple's boundary value problem solver. (b) Full
numerical simulations starting from a spot-state (\ref{ss}, \ref{w}) as
initial conditions. 
First row: spot deformation, $f_0=1.3 , D=0.35$.
Note that the final
stedy state of the system is a deformed blob shown here at $t=4000$.
Second row: self-replication, $f_0=2.2, D=0.24$. 
Third row: spot-to-ring
bifurcation, $f_0=1.3, D=0.1$. Fourth row: spot-to-ring-to-spots
bifurcation, $f_0=2.6, D=0.1$.}
\label{fig:bif}
\end{figure*}

Numerical computations indicate that self-replication is more prevalent
for spots of small radius (see Fig.~\ref{fig:bif}).
For larger spots, a deformation usually leads to 
to the so-called "finger growth'' and space filling curves. 
Others  studies have shown the occurrence of
fingering instabilities leading to labyrinthine patterns \cite{Fingers},
\cite{Goldstein}, \cite{mo}. The
question of whether the self-replication or fingering instability 
occurs first is still open.

For smaller, $O(1)$ values of $D$, there is also a different instability
mechanism that can lead to splitting of a spot into a ring as illustrated
in Fig.~\ref{fig:bif}.  Unlike spot multiplication, this instability
is radially symmetric, and is caused by the {\em dissapearence} of the
steady state solution -- whereby the steady state solution ceases to
exist due to the presence of a saddle-node bifurcation -- 
rather than by its lateral instability. Numerical simulations suggest
that that spot replication occurs only for spots of smaller radius,
whereas spot-to-ring instability is dominant for larger spots.

To conclude, 
we have estimated analytically for which
value of diffusion $D$ spot deformation first occurs. 
As is $D$ is decreased further, self-replication and/or spot-to-ring
instability is observed. As evidenced by numerical simulations, we
conjecture that spot deformation is the precursor to this phenomenon.

\end{document}